\documentclass[journal,a4paper,twoside]{IEEEtran}

\usepackage{cite}
\usepackage{url}

\usepackage{graphicx}
\usepackage{epstopdf}

\usepackage{balance}

\usepackage{enumerate}
\usepackage{mathtools}
\usepackage{amssymb}

\usepackage{algorithm2e}

\newcommand{\YEAR}{2020}
\newcommand{\VOLUME}{66}
\newcommand{\NUMBER}{1}
\newcommand{\PAGES}{193--199}
\newcommand{\RECEIVEDATE}{November 15, 2019}
\newcommand{\REVISIONDATE}{January, 2020}
\newcommand{\DOI}{DOI: 10.24425/ijet.2020.131863}  

\usepackage{fancyhdr}
\begin{document}
%

\title{A reactive algorithm for deducing nodal forwarding behavior in a multihop ad-hoc wireless network in the presence of errors}
%
%
%

\author{Karol Rydzewski, Jerzy Konorski
	\thanks{This work was funded by National Science Centre, Poland under grant No. UMO-2016/21/B/ST6/03146.}
	\thanks{K.~Rydzewski and J. Konorski are with Faculty of Electronics, Telecommunications and Informatics,  Gdańsk University of Technology, Gdańsk, Poland (e-mail: {k.rydzewski@o2.pl, jekon@eti.pg.edu.pl}).}
}

%
%

%
\markboth{}{}
%
%



\maketitle
\thispagestyle{fancy}%
\markboth{K.~Rydzewski, J.~Konorski}{A reactive algorithm for deducing nodal forwarding behavior in a multihop ad-hoc wireless network in the presence of errors}
%

 
\begin{abstract}
A novel algorithm is presented to deduce individual nodal forwarding behavior from standard end-to-end acknowledgments. The algorithm is based on a well-established mathematical method and is robust to network related errors and nodal behavior changes. The proposed solution was verified in a network simulation, during which it achieved sound results in a challenging multihop ad-hoc network environment.
\end{abstract}



\IEEEpeerreviewmaketitle


\section{Introduction}
%
%
%
%
\IEEEPARstart{S}{elf-organization} and node autonomy in ad-hoc networks pose many research and engineering challenges. One important issue is nodal selfishness and a strong microeconomic incline toward uncooperative behavior — primarily, dropping some or all offered transit packets instead of forwarding them \cite{li2008}. Many concepts aiming at incentivizing cooperative behavior have been proposed; two main lines of research in this area are micropayment schemes \cite{buttyan2002} and reputation systems \cite{hoffman2009}. The presented solution comes under the latter line, and uses a novel approach to detection of nodal forwarding behavior, the core mechanism of every reputation system. In our research, we developed two algorithms based on the same model to solve this problem. One of them, described and validated with the same simulated data set, cf. reference \cite{konorski2019}, uses linear programming enhanced with heuristics to address network errors and occasional changes in forwarding behavior. As an output it produces for each node an interval containing, in most cases, the node's true percentage of forwarded packets, thereby also providing a measure of confidence in the deduced behavior. In contrast, the algorithm detailed in this paper uses the least squares method, which is better suited to operate in environments where changes and errors are prevalent. We incorporated a mechanism in the algorithm that improves the deduction quality in volatile environments and informs users of possible deduction error.

The remainder of this article is organized as follows: Sec. II summarizes related works and highlights the advantages of our solution; Sec. III presents the adopted model; Sec. IV details the deduction algorithm; Sec. V discusses the simulation environment and the results of the algorithm's evaluation; finally, Sec. VI concludes the paper.


\section{Related work}
In recent years, many concepts of reputation systems have been proposed for multihop wireless networks to ensure that (selfishly) misbehaving nodes are isolated or forced into cooperation; see reference \cite{hoffman2009} for a survey. An essential part of such systems is a behavior detection mechanism. 

Many of these systems use a watchdog mechanism \cite{buchegger2002},\cite{michiardi2002},\cite{gupta2013},\cite{rafaei2010},\cite{chiejina2015} that relies on omnidirectional single-channel wireless transmission for sensing packets transmitted by a neighbor node. Its fundamental assumption is that if a node A requests a neighbor node B to forward its packets to a remote node C, then a watchdog at node A can also check whether node B forwards the packets to node C. However, owing to various imperfections of wireless transmission, watchdogs are known to be unreliable. Moreover, they do not reflect the nature and underlying incentives of a multihop wireless network service, which consists in source-to-destination rather than node-to-node packet transfer, thus violating the end-to-end principle \cite{saltzer1984}.

The need to derive agent reputation from a service composed of multiple agents has been pointed out in \cite{paracha2012}. However, research efforts to date have focused on the explicit locations of misbehaving nodes on paths, e.g., via the two-ACK scheme \cite{gopalakrishnan2011}, node auditing \cite{zhang2016}, or end-to-end flow conservation analysis \cite{graffi2007}.

A notable example, proposing a different approach addressing the above need, has been presented in \cite{sivanantham2013} as a a concept of selfish node detection in military wireless sensor networks with hierarchical tree structures. In this concept, the detection and later avoidance of malicious nodes is orchestrated by the sink node that periodically changes the network topology and assigns a unique encryption key to every node of the network. A source node adds a sequence number and the node’s ID to each generated packet and encrypts the packet. Every node further along the path adds its own ID to the packet and encrypts it again with its own key. A sink gathers statistics on network behavior. Given the network topology and its changes, and knowing the ratio of delivered packets to packets sent by a given source node, the sink is able to deduce malicious and suspected nodes in the network. Obviously, the system is constrained to work in controlled networks, and is unfit to work in self-organizing ad-hoc networks. Additionally, it requires an encryption component into every network node, which can result in a prohibitive cost in some low-powered devices.

Another, similar mechanism working in multihop ad hoc networks with OLSR routing protocol has also been proposed \cite{tan2016}. The algorithm collects information on all paths used within a given network, i.e., on the transit nodes as well as packets sent and delivered on each path. The data are then analyzed using a heuristic algorithm to identify misbehaving nodes. The algorithm attempts to isolate a single node that exists only on misbehaving paths and does not exist on any path working correctly, through comparing the paths’ intermediate node sets. If such a node is identified, it is blamed for the unsatisfactory performance of all paths on which it is present. As the authors acknowledge, this method has several limitations: the network must have a sufficiently large set of disjoint paths, the number of misbehaving nodes must be much smaller than the number of cooperative nodes, and only a limited number of misbehaving nodes may be detected. The algorithm is incapable of identifying any misbehaving node when a set of misbehaving paths exists such that there is no single node present on all of these paths, since then at least two nodes are responsible for this situation. Similarly, the algorithm fails to reflect reality when a single node from a set of misbehaving paths can be isolated, but it is only partially responsible for the decreased path performance, because there are more nodes contributing to it. In such cases, only one node, not necessarily the worst-behaving one, is identified and takes responsibility for all the misbehavior, whereas other misbehaving nodes are considered fully cooperative. Fine-grained numbers representing node behavior are then translated into one of three classes of nodal behavior via fuzzy-logic rules. The above mechanism can be employed to detect two types of misbehavior: dropping and delaying packets. Each of these detection schemes is based on end-to-end acknowledgment, the former on packet delivery statistics and the latter on the number of packets not delivered within an a priori defined timeframe. The type of misbehavior responsible for the resulting rating cannot be specified.

Many detection mechanisms can discern only cooperative or misbehaving nodes by measuring their behavior against a predefined desirable pattern. Most often they rely on threshold-based criteria or fuzzy-logic rules. Others \cite{konorski2010} create a finer-grained view of a node, typically with real-valued reputation levels between 0 and 1.

The mechanism proposed in reference \cite{tan2016} is similar to the solution described in this work, however, our algorithm seems more universal and robust, as demonstrated in subsequent sections. We state the major differences between our solution and the referenced works as follows:
\begin{itemize}
    \item The solution proposed in this work is able to detect an unlimited number of misbehaving nodes and deduce accurately behavior levels for each of them.
    \item The deduced behavior precisely reflects actual nodal forwarding behavior and therefore enables a finer network response.
    \item The algorithm provides additional information on the quality of deduction, which can be used to assess its usefulness in a given situation.
    \item The algorithm operates in a reactive manner and is constantly optimizing its output to mitigate the effects of nodal behavior changes and network errors.
    \item The proposed solution can work with any routing protocol, providing a source node with information on all intermediate nodes on the route used.
    \item The algorithm is based on a classical optimization problem with well-known properties and many dedicated high-performance solvers available.
    \item The algorithm works in any topology without the need for introducing any supporting infrastructure, e.g., encryption or authentication. Consequently, the algorithm does not degrade the flexibility, autonomy and performance of the network.
    \item Our solution does not rely on the assumption that transit nodes will cooperate in detecting behavior of their peers, thus it is compliant with the widely recognized end-to-end principle \cite{saltzer1984}.
\end{itemize}

\section{Model}
In our notation, sets are written in boldface, nodes are denoted by upper-case letters, and lower-case symbols are reserved for various numerical characteristics. The set of network nodes is denoted by \textbf{N}. For simplicity, in the presented solution it is a static set; however, the algorithm can work without modifications on a variable set. The traffic pattern is represented by a set \textbf{K} of source-destination node pairs. If there is no wireless link between a given source-destination node pair, then a path involving transit nodes is established by using some single-path routing protocol. For a path \textit{k} $\in$ \textbf{K}, let \emph{S$_k$}, \emph{D$_k$} $\subseteq$ \textbf{N} denote the source and destination nodes, and let $\textbf{X}_k \subseteq \textbf{N} \setminus \{S_k,D_k\}$ denote the (possibly empty) set of transit nodes\footnote[1]{The sequence of nodes in \textbf{X$_k$} is irrelevant as long as the nodes can form a valid path between \emph{S$_k$} and \emph{D$_k$}.}. 

As previously indicated, the transit nodes may selfishly drop all or part of the offered transit packets. To incentivize a satisfactory packet forwarding service, the network must first accurately deduce nodal forwarding behavior. To this end, it incorporates a reputation system whose task is to deduce each node's forwarding behavior from observable performance characteristics. Later use of the deduced behavior levels is beyond the scope of this work. Although this task could be fulfilled in a distributed fashion, for ease of exposition we assign it to a single network-wide \emph{reputation server} (RS). (A distributed algorithm will be investigated in future work.) Source nodes in the network, upon completion of each communication session, send a report on the observed end-to-end packet delivery ratio (PDR). After receiving the report from a source node, RS runs a dedicated algorithm that calculates each node's forwarding behavior level. Hence, watchdogs are dispensed with, and deducing the individual forwarding behavior of the nodes in $\textbf{X}_k$ from the PDR on path \emph{k} becomes the main challenge.

With regard to the forwarding behavior of a node $\emph{X} \in \textbf{N}$, we introduce two quantities. One, denoted by $g_X$, is its \emph{intrinsic forwarding trustworthiness} (IFT) defined as the percentage of offered transit packets that the node is inclined to forward toward the destination. This is a ground truth-type quantity, which is accurately observable only to node \emph{X} itself. The other quantity, denoted by $d_X$, may change as new reports are received; it is the nodal behavior level as deduced by RS from the reported PDRs. We have $g_X, d_X \in [0, 1]$, where 0 signifies a lack of cooperation (no packet forwarding), and 1 signifies fully cooperative behavior (no packet dropping). Ideally, $d_X = g_X$, but in reality the two quantities may differ on account of inaccurate observation of end-to-end PDRs and possible ambiguities produced by the underlying deduction algorithm at RS. Accurately approximating $g_X$ by $d_X$ is critical to the viability of our reputation system.

The proposed deduction algorithm relies on the following assumptions:
\begin{enumerate}[i]
\item After each packet transmission (e.g., TCP) session on path $k \in$ \textbf{K}, the source node $S_k$ calculates and truthfully reports to RS the observed end-to-end PDR, denoted by $p_k$, along with the set \textbf{X}$_k$. The presence of \textbf{X}$_k$ in the report implies that the employed routing protocol reveals all transit nodes on path \emph{k} to $S_k$ before all packets within the session have been transmitted toward $D_k$. An example is Dynamic Source Routing (DSR) \cite{johnson2001}, an on-demand source routing protocol in which $S_k$ initiates a path discovery process, and one path is returned in a reply message containing a list of all discovered transit nodes. This list is carried in packets' DSR headers (and is also visible to all the nodes in \textbf{X}$_k$).
\item Node \emph{X} forwarding behavior with respect to an offered transit packet is defined as an IFT-based forward/drop decision, whereby a retained packet is forwarded with probability $g_X$ and dropped with probability $1-g_X$.
\item The forward/drop decisions at the nodes in \textbf{X}$_k$ are fully autonomous and statistically independent, and also are not selective with regard to \emph{k}. Path selective behavior is not assumed to occur, for simplicity of analysis, although such behavior is generally possible \cite{tan2016} and will be investigated in future work. External, e.g., temporary congestion-related factors may influence the observed PDRs, thus possibly resulting in apparent path selective behavior; in our model, these factors are regarded as observation imperfections and are not explicitly modeled.
\end{enumerate}

In light of assumptions i–iii, the probability of successful packet delivery on path \emph{k}, as observed by $S_k$ through the analysis of incoming end-to-end ACKs and later reported to RS as $p_k$, is determined by the path equation:
\begin{equation} \label{eq:1}
p_k=\begin{cases}
    {\prod \limits_{X\in \textbf{X}_k}} g_X,&\textbf{X}_k\neq\varnothing \\
    1,&\textbf{X}_k=\varnothing
\end{cases}
\end{equation}

\begin{figure}[h]
   
	\begin{center}
		\includegraphics[width=1.0\linewidth]{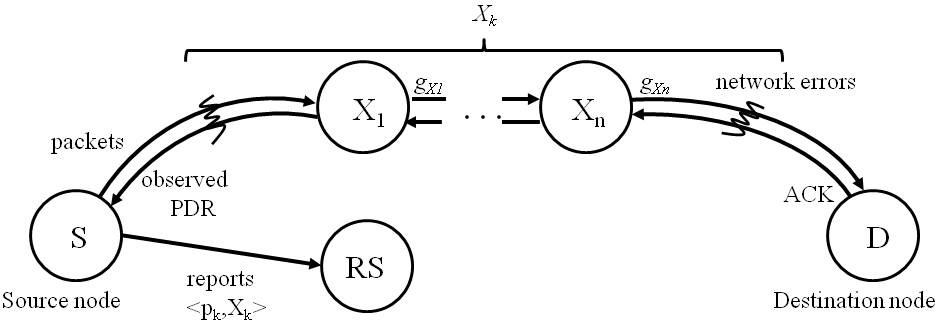} 
	\end{center}
	\caption{Path model and communication with RS.}\label{fig:fig1} 
\end{figure}

Note that $p_k$ is the fully determined ground-truth IFT of the nodes in \textbf{X}$_k$. If path \emph{k} contains only ideally trustworthy transit nodes (or no transit nodes at all), then $S_k$ can expect an end-to-end PDR equal to 1. If path \emph{k} does not contain any transit nodes (\textbf{X}$_k = \varnothing$), $p_k = 1$, as $S_k$ is clearly interested in forwarding all of its source packets. A similar model of $p_k$ in a selfish network environment has been proposed in \cite{liu2010}. The path model, RS communication and nodal forwarding behaviors are illustrated in Fig.~\ref{fig:fig1}.

The centralized RS model simplifies the considerations, although a distributed reputation system can be envisaged instead; this will be examined in future work. In addition, we stress that the proposed deduction algorithm has reasonable routing requirements (information on transit nodes available for the source node prior to a session end) and can work with any number of misbehaving nodes.

Nodal behavior levels are calculated by RS whenever a PDR report is received, on the basis of this and all previously received PDR reports. In the next section, we first describe the algorithm for an idealized model, and next proceed with mechanisms for mitigating PDR observation errors, IFT changes and possibly untruthful reporting.

\section{Algorithm}

Each source node in the network collects data on the PDR ratio ($p_k$) and transit nodes (\textbf{X}$_k$) for each path (\emph{k}) that the node uses. The data tuple $\langle p_k, |\textbf{X}_k|\rangle$ is sent to the RS upon completing a TCP session on a given path; this tuple constitutes a PDR report. The RS builds an equation system from all received reports according to the model (\ref{eq:1}):

\begin{equation} \label{eq:2}
p_k={\prod \limits_{Y\in \textbf{X}_k}} g_Y,\forall k \in \textbf{K}
\end{equation}

Because a multiplicative PDR model equation is not suitable for linear regression methods, the RS uses logarithmic transformation to obtain a system of linear equations. Let $\widetilde{p}_k = -\log_b(p_k)$ and $\widetilde{g}_X = -\log_b(g_X)$, the logarithms being to any base $b > 1$. Then (\ref{eq:2}) transforms into:

\begin{equation} \label{eq:3}
\widetilde{p}_k={\sum \limits_{Y\in \textbf{X}_k}} \widetilde{g}_Y,\forall k \in \textbf{K}
\end{equation}

The linear system (\ref{eq:3}) is an input for a behavior deduction algorithm, which is identical to a least-squares minimization problem:

\begin{equation} \label{eq:4}
\text{minimize: } |A\widetilde{g}-\widetilde{p}|
\end{equation}
\begin{equation}
\text{subject to: } g_Y \in [0,1], Y\in\textbf{N}\nonumber
\end{equation}
where \emph{A} is an $|\textbf{N}|\times|\textbf{K}|$ node-path incidence matrix, and   is a $|\textbf{K}|$-dimensional column vector with a generic entry $\widetilde{p}_k$. Writing $A^T$ for the transpose of \emph{A}, the solution to (\ref{eq:4}) can be routinely obtained as:

\begin{equation}\label{eq:5}
    \widetilde{g}^*=(A^TA)^{-1}A^T\widetilde{p}
\end{equation}
and subsequently transformed to the deduced behavior level $d_X, X \in \textbf{N}$, using  $d_X=\exp_b(-\widetilde{g}_X^*)$. The result is a scalar number representing the best possible match of all $d_X$ for all equations in the RS-built linear system, i.e., $d_X$ minimizes the total discrepancy among $p_k$ (PDR observed by source nodes) and $p_k^{'}$ (PDR calculated by substituting $d_X$ into all the equations in the system). This discrepancy is called the \emph{residual}. In the basic least-squares method, the error of deduction of $d_X$ follows from (\ref{eq:4}). This error may be of fundamental significance if $d_X$ is used as a decision factor for subsequent cooperation enforcement. To give this error, denoted $e_X$, a more straightforward interpretation, we propose a method to assess it as the maximum residual for all reported paths containing node \emph{X}:


\begin{equation} \label{eq:6}
   e_X=\begin{cases}
   \max_{\substack{k\in\textbf{K}_X}}|p_k-p_k^{'}|,&|\textbf{K}_X|>3\\
   penalty,&|\textbf{K}_X|\le3
   \end{cases}
\end{equation}
where $p_k$ is given by (\ref{eq:2}), $p_k^{'}={\prod \limits_{Y\in \textbf{X}_k}}d_Y$, and $\textbf{K}_X=\{k \in \textbf{K}: X \in \textbf{X}_k\}$. 
When the number of reported paths comprising node \emph{X} as a transit node is small, the $e_X$ value is set to an a priori defined \emph{penalty} value. The penalty is set to 1 (highest possible error value) to avoid yielding misleading results when the number of reports on \emph{X} behavior is insufficient to yield a conclusive estimate. Without this provision, algorithm (\ref{eq:6}) would significantly underestimate the $e_X$ for such nodes. The threshold value  3 was experimentally evaluated to provide a satisfactory balance between an acceptable risk of a false result and the minimum sample size required to yield informative output. 

The formula (\ref{eq:6}) produces a single number that can be used to calculate an interval within which $g_X$ should be located: $g_X \in [d_X-e_X,  d_X+e_X]\cap[0,1]$. Importantly, the interval has the following properties:
\begin{enumerate}
    \item It is based on source nodes’ reports; in some cases, the true $g_X$ may be located outside this interval. An example of such a case is a situation where majority of \textbf{K}$_X$ reports experience unusually high medium-related losses. This may happen if a relatively small portion of a network suffers from congestive traffic, while the rest of the network operates in more favourable conditions. The algorithm should gradually recover as new reports are received.
    \item The interval depends on errors related to all nodes coexisting with \emph{X} at least once on a path in \textbf{K}$_X$. It is impossible to deduce which node is responsible for a specific value of $e_X$.
\end{enumerate}
One path in \textbf{K}$_X$ almost always resulted in a large ($|g_X-d_X+e_X| > 0.05$) deduction error for $X$ because of  interference between the PDR-reading error and the inherent least-squares error. Two paths in \textbf{K}$_X$ improved the results, however, significant errors were observed in roughly $10\%$ of the cases. With $|\textbf{K}_X|=3$ we did not observe any significant errors and even small errors ($|g_X-d_X+e_X| < 0.05$) were infrequent. Note that because the results of the deduction algorithm strongly depend on the network topology and the nodes' IFT, it is hard to judge if the adopted threshold value is a universally optimal choice.

In typical circumstances (without nodal behavior changes and under steady traffic conditions), the value of the observed error initially increases with incoming PDR reports and eventually stabilizes. Unusually large network errors (e.g., buffer overflow, transmission impairments, MAC-layer delays, end-to-end acknowledgment loss, PDR inaccuraccies) and nodal IFT changes result in sharp surges in $e_X$. The value by which $g_X$ changes is correlated with the increase in $e_X$. The presented solution uses $e_X$ values to improve the \emph{deduction accuracy} (defined as $g_X-d_X$) and enable prompt response to nodal IFT changes.

Proposed algorithm operates accordingly to Algorithm \ref{alg}. Upon receipt of a new report, RS transforms it with logarithm transformation (\ref{eq:3}), and calculates $d_X$ and $e_X$ values for each node in \textbf{N}, based on the whole \textbf{K}, including the newly received report. Then, RS starts search for possible optimization by looking for a \textbf{K}$_X$ subset, which removal results in lowering overall $e_X$. Search for improvement is continued until no suitable \textbf{K}$_X$ is identified for removal or further search is impossible (e.g., all paths have been removed).

\begin{algorithm}
\caption{A reactive algorithm for deducing IFT}
\label{alg}
\Begin(\text{upon reception of a PDR report: $<p_k$,\textbf{X}$_k>$})
{
$<p_k$,\textbf{X}$_k>$ \emph{report} $\gets$ log. trans.($<p_k$,\textbf{X}$_k>$); //(\ref{eq:3})\\
\While{true}{
vector \emph{orig.} $d_X \gets$ deduce IFT(\textbf{K}$ \cup$ \emph{report}); //(\ref{eq:5})\\
vector \emph{orig.} $e_X\gets$ estimate error (\textbf{N}, \emph{orig.} $d_X$, \textbf{K} $\cup$ \emph{report}, penalty); //(\ref{eq:6}) for each $X\in$ \textbf{N}\\
\ForEach{$X \in$ report.\textbf{\upshape X}$_k$}
{
	vector $d_X \gets$ deduce IFT((\textbf{K}$\setminus\textbf{K}_X$) $\cup$ \emph{report})\;
	$e_X \gets$ sum(estimate error(\textbf{N}, $d_X$, (\textbf{K}$\setminus\textbf{K}_X$) $\cup$ \emph{report}, penalty))\;
}
$<node,e_X> best\gets$ min($e_X$)\;
\uIf{$best.e_X \geq$ sum(\emph{orig.} $e_X$)}
{
	\textbf{K} $\gets$ \textbf{K} $\cup$ \emph{report}\;
	orig. $e_X \gets$ estimate error(\textbf{N}, orig. $d_X$, \textbf{K}, \emph{penalty}=1) 
	\Return \emph{orig.} $d_X$, \emph{orig.} $e_X$\;
}
\Else{
	$\textbf{K} \gets \textbf{K}\setminus\textbf{K}_{best.node}$\;
}
}
}
\end{algorithm}
Notably, removing more paths from \textbf{K} than necessary is undesirable because it negatively affects deduction for all nodes in the network until new reports are received. Because of this fact, excessive path removal should be discouraged. Assigning an appropriate penalty value \ref{eq:6} is essential for achieving a good balance between the sensitivity of the deduction algorithm for IFT changes and the prevention of unnecessary path removal. A discussion of the influence of the penalty values on the algorithm's deduction accuracy will be provided later. Note that in Algorithm \ref{alg} there is a difference between the penalty values returned by the algorithm as an end-result, to provide information on possible deduction errors (penalty always equals 1), and the penalty value used to decide weather to remove \textbf{K}$_X$ subset (penalty $\in[0,1]$).

Another measure implemented in the presented solution restricts the maximum size of the analyzed linear system to some predefined value called \emph{history}, denoted by \emph{h}. This reduces the resources (memory and computing power) required for deduction, and ensures that PDR reports produced by outdated nodal IFT are eventually removed even if the path removal mechanism based on $e_X$ differences fails to correctly identify such reports. The \emph{h} must be large enough to ensure minimal disruption when paths are removed through the $e_X$-based mechanism. In stable environments, longer histories guarantee superior accuracy; in contrast, in environments where nodal IFT changes are frequent, decreasing \emph{h} is beneficial as it speeds up response to nodal IFT changes and permits to quickly return to acceptable accuracy, cf. experimental results in Sec. V.

\section{Simulations}

The simulation environment was built in OMNeT++ with the Inetmanet extension package \cite{omnet}. A simulated network was configured as a multihop ad-hoc wireless network based on the IEEE 802.11g standard. Selected simulation parameters are provided in Table \ref{table}.

\begin{table}[h] 
	\renewcommand{\arraystretch}{1.3}
	\caption{Selected parameters of the simulator setup.}
	\label{table}
	\centering
	\begin{tabular}{|c|c|}
		\hline
		\textbf{Parameter} & \textbf{Value}\\
		\hline
		antenna & omnidirectional\\
		\hline
		nodal transmission & 1 mW\\
		power & \\
		\hline
		receiver sensitivity & -90 dBm\\
		\hline
	    transmission error & Ieee80211BerTableErrorModel\\ 
	    model& (''per\_table\_80211g\_Trivellato\.dat'')\cite{omnet} \\
		\hline
    	MAC protocol & 9 Mbps 802.11g\\
		\hline
		EDCA & enabled \\
		\hline
		maximum queue size & 50 \\
		\hline
        network layer ACK & disabled \\
		\hline
		routing protocol & DSR, route request period = 1 s\\
		\hline
		transport protocols & TCP \\&UDP (only for to- and from-RS messages) \\
		\hline
		TCP mode & DumbTCP \cite{omnet} \\
		\hline
		TCP settings & disabled: delayed ACK, selective ACK, \\&Nagle’s algorithm \cite{nagle1984}; \\&maximum segment size = 1452 B, \\&advertised window = 65535 \\
		\hline
		simultaneously active  & 1-4\\TCP sessions & \\
		\hline
		nodal behavior & 0.5-1.0\\
		\hline
	\end{tabular}
\end{table}

\begin{figure}[h]
	\begin{center}
		\includegraphics[width=1.0\linewidth]{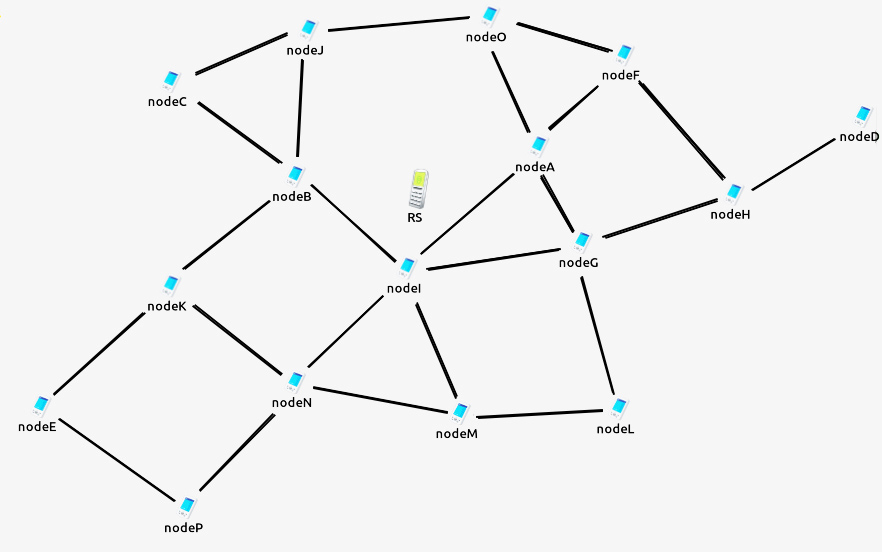} 
	\end{center}
	\caption{Network used in simulations (wireless links marked with solid lines).}\label{fig:fig3} 
\end{figure}

Node locations, as depicted in Fig.~\ref{fig:fig3} along with the available wireless links, were invariable throughout the simulations. However, source-destination pairs were randomly selected for each path (thus implying randomly selected transit nodes). Moreover, initial nodal IFTs were randomly assigned at simulation startup. Hence, the examined network topology was somewhat different in every simulator run. Additionally, other networks, with varying numbers (up to 15) and location of the nodes, were examined in parallel to this research and similar results were achieved. Multiple traffic patterns were examined, in which the average packet loss ratio due to network error (mostly buffer overflow) fluctuated below 5\%, with sporadic spikes up to 11–12\%.

Nodal IFTs varied according to a stochastic model. An input to this model was the mean number of initiated sessions between singular behavior changes in the entire population of nodes, denoted by $\tau$. On the basis of $\tau$ and the total number of nodes in the network, $|\textbf{N}|$, the probability of IFT change for a single node ($c_g$) per initiated session was calculated as:

\begin{equation} \label{eq:7}
    c_g=\frac{1}{\tau|\textbf{N}|}
\end{equation}

Subject to the above stochastic model, changes in a node's IFT were simulated when the node was chosen to be a transit node on a newly created path on condition that it was not currently serving as a transit node on another active path. We chose such a model to simplify the control and analysis of nodal behavior changes. Alternatively, the behavior could change during an active session, resulting in the effective nodal behavior during that session being some combination of the initial and the new behavior. (An extreme option would be for a node to behave differently during any two different sessions; we conjecture that the algorithm should be able to perform correctly under such a model as well, however, because of the additional complexity of configuring the algorithm properly, we defer this to future work.) The new behavior of a node was stochastically drawn according to a uniform distribution within the assumed domain for the simulations ($g_X\in[0.5,1]$). Six $\tau$ values were used: $\{10, 50, 100, 150, 200 \text{ and } \infty\}$, with multiple simulation runs performed for each. The presented algorithm for behavior deduction and observed error evaluation was examined on uniformly prepared traffic generation data. The influence of the penalty and history parameters was evaluated and compared with the standard least-squares method not featuring these parameters (\ref{eq:5}). A total of 483 combinations of penalty and history parameters were examined, where $penalty \in \{0, 0.05, 0.1, 0.15, …, 0.95,1\}$ and $h \in \{1, 25, 50, 75, …, 475, 550\}$.

\begin{figure}[h]
	\begin{center}
		\includegraphics[width=0.8\linewidth]{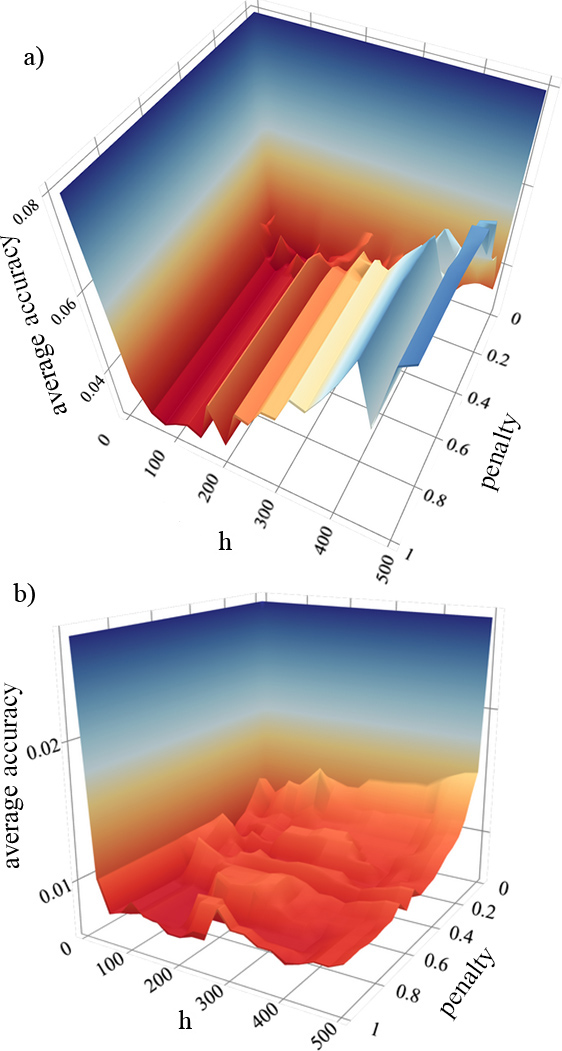} 
	\end{center}
	\caption{Average absolute accuracy in relation to penalty and history settings in the deduction algorithm, (a) $\tau = 10$, (b) $\tau = 100$; for clarity, values above 0.08 in (a) and 0.028 in (b) were reduced to these values.} \label{fig:fig4}
\end{figure}

Fig.~\ref{fig:fig4} presents the average absolute value of deduction accuracy ($|g_X-d_X|$) in two simulation runs with different $\tau$ values ($\tau = 10$ in Fig.~\ref{fig:fig3}a and $\tau = 100$ in Fig.~\ref{fig:fig3}b in relation to variable $h$ and \emph{penalty} parameters (exploring their full spectrum studied in the experiment). Steep spikes appear for low values of either: insufficient numbers of reports (low \emph{h}) render accurate deduction impossible and result in high $e_X$ rates; similarly, low \emph{penalty} values do not prevent excessive path removal, and the deduction algorithm is fed with an insufficient number of path reports. The shape of the surface beyond these extremely low values depends on the frequency of changes. For low $\tau$ values, the deduction accuracy achieves the lowest values for \emph{penalty} in the lower part of its spectrum, e.g., \emph{penalty} $\in(0.1, 0.4)$ for $\tau = 10$. The \emph{h} values have limited influence; generally, \emph{h} in the upper end of the tested range yield the smallest deduction accuracy for comparable \emph{penalty}; however, in environments where IFT changes are prevalent, a short history may yield better results. The lower the IFT change frequency (i.e., the larger $\tau$), the more level the surface become; in such stable environments, high values of both penalty and history yield the best results.

\begin{figure}[h]
	\begin{center}
		\includegraphics[width=0.9\linewidth]{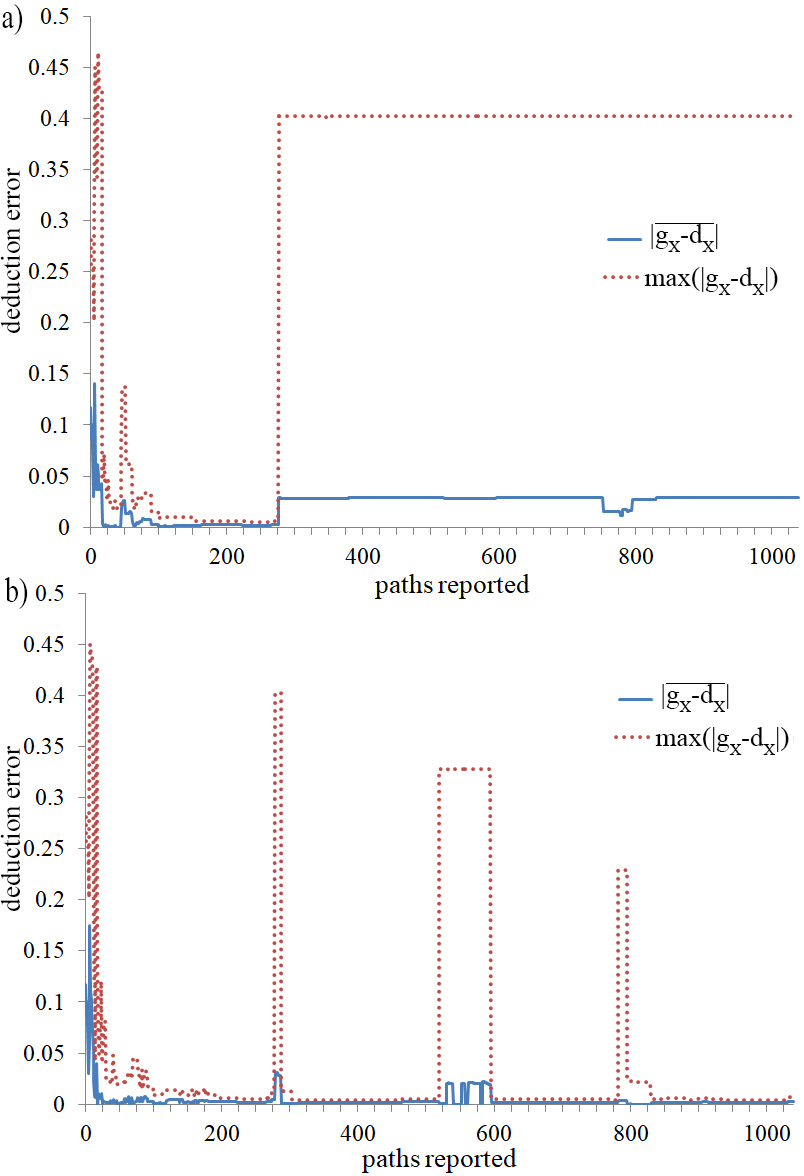} 
	\end{center}
	\caption{Outcome of the least-squares algorithm, (a) the simple variant, (b) enhanced variant with path removal and limited history; solid lines: average $|g_X-d_X|$ for all the nodes, dotted lines: maximum $|g_X-d_X|$.} \label{fig:fig5}
\end{figure}

Fig.~\ref{fig:fig5} compares the results achieved by the deduction algorithm with and without its error-reacting component (\ref{eq:6}) (respectively, Fig.~\ref{fig:fig5}b and Fig.~\ref{fig:fig5}a). In both plots, two values are plotted in relation to the number of consecutively received reports: average $|g_X-d_X|$ (solid line) and maximum $|g_X-d_X|$ (dotted line). The outcome in Fig.~\ref{fig:fig5}b was achieved with \emph{penalty} = 0.85 and $h = 325$, which are the highest \emph{penalty} and $h$ values among a group of the lowest $|g_X-d_X|$ results observed in Fig.~\ref{fig:fig4}b. The nodal IFT change frequency was $\tau = 100$. The IFT changes and significant network errors were quickly detected, and the linear system at RS was aptly trimmed. Acceptable average deduction accuracy was kept, and the most erroneous $d_X$ values were quickly fixed. In contrast, the standard least-squares algorithm was incapable of adapting to nodal IFT changes and keeping the deduction accurate. 

The simulation results demonstrate that report removal to minimize $e_X$ and history restriction are effective means of improving behavior deduction accuracy when changes in IFT behavior are prevalent. It was demonstrated that by manipulating the \emph{penalty} and $h$ parameters, decreasing in a volatile environment and increasing in a stable environment, the algorithm performance can be adjusted to produce acceptable results and short response times under various traffic conditions.

\section{Conclusion}

A novel algorithm for deducing nodal individual forwarding trustworthiness (IFT) on the basis of end-to-end acknowledgements was presented. The algorithm is based on a well-established mathematical method and incorporates apparatus enhancing its deduction performance in volatile network environments, as well as offers additional information on the deduction error. As such, it offers satisfactory situational awareness and live insight into algorithm's performance.During extensive network simulations, the solution was demonstrated to be robust to network-related errors as well as nodal IFT changes, and to achieve significantly better results than existing algorithms.

A decentralized deduction algorithm, working with other types of routing algorithms and better suited for mobile networks, and a scheme to incentivize cooperative nodal behavior based on the presented solution are avenues of near-future research.

\balance






\end{document}